\documentclass[%
 reprint,
superscriptaddress,
 amsmath,amssymb,
aps,
]{revtex4-2}

\usepackage{graphicx}
\usepackage{dcolumn}
\usepackage{bm}

\usepackage[hypertex,colorlinks=true,linkcolor=black,citecolor=blue,filecolor=blue,urlcolor=blue,setpagesize=false,nesting=true]{hyperref}

\begin{document}

\preprint{APS/123-QED}

\title{Antiferromagnetic structure of alkali metal superoxide CsO$_2$}

\author{Takehito Nakano}
\email{takehito.nakano.phys@vc.ibaraki.ac.jp}
\affiliation{Institute of Quantum Beam Science, Graduate School of Science and Engineering, Ibaraki University, Mito, Ibaraki 310-8512, Japan}
\author{Shun Kontani}
\affiliation{Institute of Quantum Beam Science, Graduate School of Science and Engineering, Ibaraki University, Mito, Ibaraki 310-8512, Japan}
\author{Masatoshi Hiraishi}
\affiliation{Institute of Quantum Beam Science, Graduate School of Science and Engineering, Ibaraki University, Mito, Ibaraki 310-8512, Japan}
\author{Kaito Mita}
\affiliation{Department of Physics, Okayama University, Okayama 700-8530, Japan}
\author{Mizuki Miyajima}
\affiliation{Department of Physics, Okayama University, Okayama 700-8530, Japan}
\author{Takashi Kambe}
\affiliation{Department of Physics, Okayama University, Okayama 700-8530, Japan}

\date{\today}

\begin{abstract}
We have performed a powder neutron diffraction study on CsO$_2$, where the unpaired electron with $s=1/2$ in the $\pi^*$ orbital of the O$_2^-$ ion is responsible for the magnetism. The magnetic reflections 0~$\frac{1}{2}$~0 and 0~$\frac{1}{2}$~1 were observed below the N\'{e}el temperature of about 10 K. An antiferromagnetic structure with a propagation vector of (0~,$\frac{1}{2}$,~0) and magnetic moments parallel to the $a$-axis is the most plausible. The magnitude of the ordered moment is about 0.2 $\mu_B$, which is considered to be strongly suppressed due to the one-dimensionality of the system. We propose a possible $\pi^*$ orbital order that can explain the obtained magnetic structure, and discuss its relation to the one-dimensionality. 
\end{abstract}

\maketitle

In correlated electron systems, the cooperative coupling of electronic spin and orbital produces a variety of fascinating phenomena, and many such studies have been carried out on transition metal compounds \cite{Kugel,Tokura}.
Also in some molecular solids, the spins are strongly coupled to the lattice and orbital degrees of freedom, and they can be good playgrounds to study various cooperative quantum phenomena \cite{NH3K3C60_2}.
In alkali metal superoxides, $A$O$_2$ ($A$ = Na, K, Rb, Cs), the doubly degenerate $\pi^*$ orbital of the O$^{2-}$ molecule accommodates three electrons (or one hole) creating a spin $s$ = 1/2 state.
A wide variety of physical properties are observed depending on the type of alkali metal \cite{Zumsteg}; spin gap states in NaO$_2$ \cite{Miyajima_PRB,Miyajima_D}, antiferromagnetic ordering in KO$_2$ \cite{KO2_Neutron} and RbO$_2$ \cite{Fahmi_JPSJ}, and one-dimensional (1D) antiferromagnetic states in CsO$_2$ \cite{Riyadi,CsO2_TLL,Miyajima_JPSJ}.

At room temperature, the CsO$_2$ crystal has a body-centered tetragonal structure with the space group of $I4/mmm$.
The Cs$^+$ ions are located at the apex and body center of the unit cell, while the O$^{2-}$ ions are located at the midpoint on the $c$-axis and at the face center of the base.
The O$^{2-}$ ions are aligned along the $c$-axis with an isotropic thermal libration. The degeneracy of the $\pi^*$ orbital is dynamically preserved in the tetragonal phase. Upon cooling, a two-step structural phase transition occurs at 150 K and at 70 K \cite{Riyadi,Miyajima_JPSJ}. The crystal symmetry of both phases appears to be orthorhombic.
The magnetic susceptibility shows a broad maximum at low temperature which can be well fitted by the Bonner-Fisher model, namely, a 1D Heisenberg antiferromagnet with $s$ = 1/2 \cite{Riyadi}. Furthermore, NMR studies suggest a spin Tomonaga-Luttinger liquid state \cite{CsO2_TLL}. A Density functional theory (DFT) calculation predicts that the symmetry lowering is induced by a cooperative (static) tilting of the O$^{2-}$ ions \cite{Riyadi}.
This can also be seen as a cooperative Jahn-Teller distortion accompanied by $\pi^*$ orbital ordering. In this structure, the 1D zigzag chain along the $b$-axis is expected to realize the 1D antiferromagnetic exchange pathway.
Below the N\'{e}el temperature of $T_N=$ 9.6 K, an antiferromagnetic ordering occurs \cite{CsO2_TLL,Fahmi_D}.
However, the high-field magnetization remains a feature of the low-dimensional magnetic system \cite{Miyajima_JPSJ}. The antiferromagnetic structure has not been experimentally resolved.
Such a rich variety of physical properties is realized through the interaction of spin, lattice and orbital degrees of freedom by using very simple chemical compositions in CsO$_2$. As an aid to understanding these mechanisms, this study aims to reveal the ground state magnetic structure of CsO$_2$ by neutron diffraction (ND).

Polycrystalline CsO$_2$ sample was synthesized using the liquid ammonia method, which is described elsewhere \cite{Miyajima_JPSJ}. For ND experiments, a powder sample of 2.3 g was placed in a vanadium-film cylinder, 10-mm diameter, and sealed in an aluminum can filled with He gas. The powder ND measurements were performed by using the HQR spectrometer at the T1-1 beamline of the JRR-3 reactor at the Japan Atomic Energy Agency. Pyrolytic graphite (PG) crystals were used for the monochromator and the analyzer. The energy of the incident neutron was 13.5 meV (wavelength $\lambda$ = 2.46 \AA). A PG filter was installed in front of the sample to remove higher-order contaminations. Horizontal collimation of the incident and scattered beam was 40 min. The sample was cooled down by using a Gifford-McMahon (GM) refrigerator.

\begin{figure*}
\includegraphics[width=15.0cm]{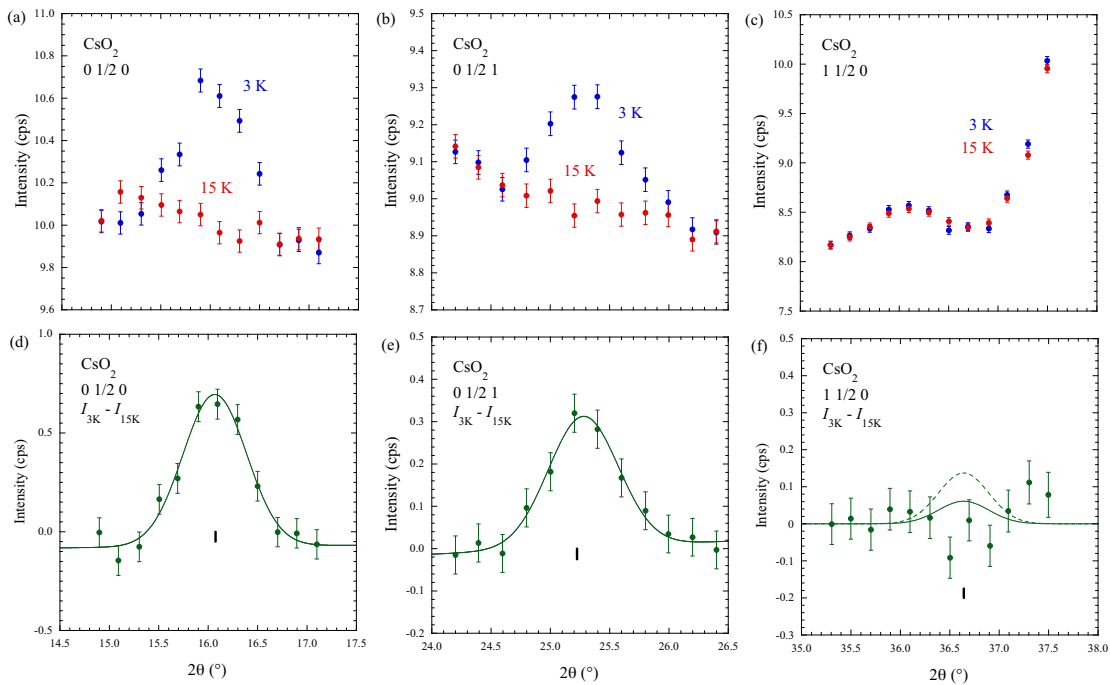}
\caption{\label{fig:magpeaks}Neutron diffraction profiles of CsO$_2$ measured at 3 and 15 K for the scattering angle ranges of (a) 0~$\frac{1}{2}$~0, (b) 0~$\frac{1}{2}$~1, and (c) 1~$\frac{1}{2}$~0 reflections. (d), (e) , and (f) are the difference plots between the data taken at 3 and 15 K. The black vertical bars indicate the expected positions of the reflections from the lattice constants. The solid curves in (d) and (e) represent the best fits using Gaussian functions. The solid and dashed curves in (f) are simulated profiles based on certain assumptions (see text).}
\end{figure*}

We measured several nuclear Bragg reflections at 15 K and confirmed that
they were consistent with the low temperature crystal structure reported by powder x-ray diffraction \cite{Miyajima_JPSJ,Miyajima_D}. The results are shown in Supplemental Material \cite{suppl}.

\begin{figure}
\includegraphics[width=0.70\linewidth]{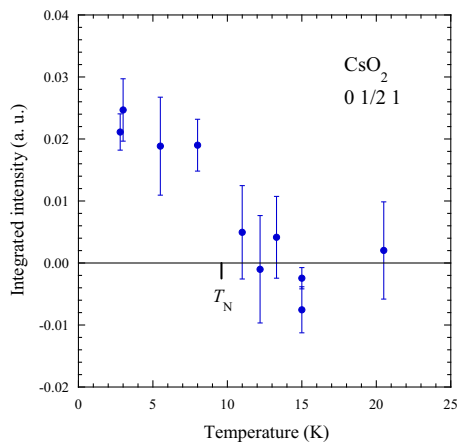}
\caption{\label{fig:orderparameter}Temperature dependence of the integrated intensity of the 0$\frac{1}{2}$1 reflection. $T_N$ obtained from magnetic susceptibility measurements is shown.}
\end{figure}

KO$_2$ is the only alkali metal superoxide whose magnetic structure has been solved \cite{KO2_Neutron}. The magnetic moments are parallel to the $ab$-plane, ferromagnetically aligned in the plane and antiparallel between adjacent planes. In this antiferromagnetic structure, the 001 magnetic reflection should be strongly observed. We measured the 001 reflection of CsO$_2$ at 3 K and 15 K, i.e. below and above $T_N$, 
and observed no growth of the magnetic reflection. Therefore, the magnetic structure of CsO$_2$ is different from that of KO$_2$.
On the other hand, the DFT calculations of CsO$_2$ by Riyadi \textit{et al}. show an expected magnetic structure with a doubling of the $b$-axis period accompanied with an orbital order. We therefore measured several reflections with the Miller index of $h~\frac{1}{2}~l$. The results are shown in Fig.~\ref{fig:magpeaks}. The upper panels show the neutron diffraction profiles measured at 3 and 15 K for the scattering angle ranges of (a) 0~$\frac{1}{2}$~0, (b) 0~$\frac{1}{2}$~1, and (c) 1~$\frac{1}{2}$~0 reflections. The lower panels (d), (e), and (f) show the difference plots between the data taken at 3 and 15 K. The 0~$\frac{1}{2}$~0 and the 0~$\frac{1}{2}$~1 reflections clearly appear at low temperatures. Their intensity is 1\% or less of the nuclear reflection intensity.
The 1~$\frac{1}{2}$~0 reflection is discussed later. The 0~$\frac{1}{2}$~1 reflection was measured at several temperatures and the temperature dependence of the integrated intensity, obtained by fitting with a Gaussian function, is shown in Fig.~\ref{fig:orderparameter}. The intensity appears below about 10 K, close to $T_N$, and increases with decreasing temperature. It is therefore a magnetic reflection. It is also natural to consider the 0~$\frac{1}{2}$~0 reflection in Fig.~\ref{fig:magpeaks}(d) as a magnetic reflection. In Fig.~\ref{fig:magpeaks}(c), there is a temperature independent peak around $36.1^\circ$. This is the 100 reflection of an impurity, CsOH$\cdot$H$_2$O \cite{suppl}. In addition, the intensity markedly increases towards the higher angles. This is due to the very intense 011 nuclear reflection of CsO$_2$ centered at $38.1^\circ$, the tail of which is observed. As seen in Fig.~\ref{fig:magpeaks}(f), there appears no peak at the expected position of the 1~$\frac{1}{2}$~0 reflection. The solid and dashed curves are simulations based on certain assumptions that will be discussed later. We have observed at least two magnetic reflections 0~$\frac{1}{2}$~0 and 0~$\frac{1}{2}$~1. From this we can conclude that the propagation vector of the antiferromagnetism of CsO$_2$ is (0, $\frac{1}{2}$, 0). This result is similar to the prediction of the DFT calculation \cite{Riyadi}. The peak widths of the two magnetic reflections shown in Fig.~\ref{fig:magpeaks}(d) and (e) are comparable to those of the nuclear Bragg peaks.
Therefore, the correlation length of the magnetic order can be regarded as long as that of the crystal structure.

\begin{figure}
\includegraphics[width=0.75\linewidth]{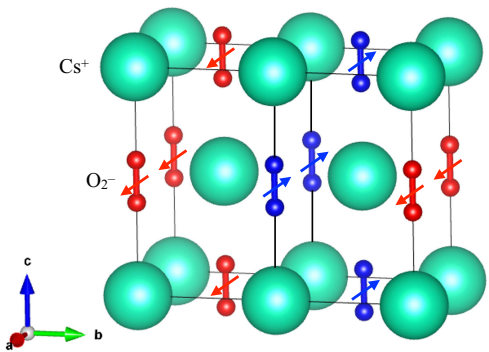}
\caption{\label{fig:MagStrct}The most plausible magnetic structure of CsO$_2$. The red and blue arrows represent the ordered magnetic moments of O$^{2-}$ ions. The propagation vector of the antiferromagnetism is (0, $\frac{1}{2}$, 0), and the 0.2 $\mu_B$ ordered moment is oriented along the $a$-axis. VESTA is used to draw the crystal structure \cite{VESTA}, which is then modified.}
\end{figure}

\begin{figure}
\includegraphics[width=\linewidth]{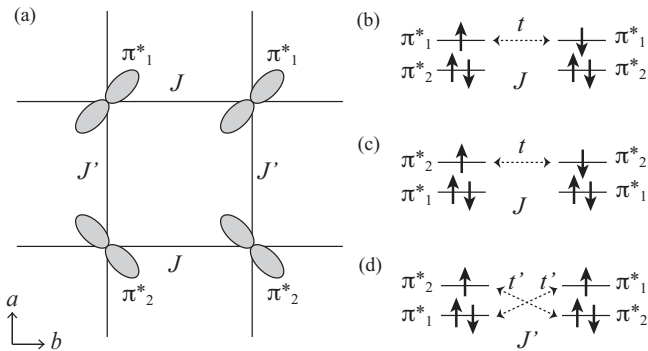}
\caption{\label{fig:orbital}A model of orbital order in CsO$_2$. (a) Schematics of the degeneracy lifted $\pi^*$ orbitals, which accommodate unpaired electrons in the $ab$-plane. A ferro-orbital order is realized along the $b$-axis, while an antiferro-orbital order is realized along the $a$-axis. (b), (c) Schematics of the antiferromagnetic exchange coupling along the $b$-axis. (d) Schematics of the ferromagnetic exchange coupling along the $a$-axis.}
\end{figure}

Here we attempt to estimate the magnitude of the ordered magnetic moment.
By comparing the integrated intensity of the 002 nuclear reflection with that of the magnetic reflection, the magnitude of the ordered moment $\mu_{khl}$ was determined using the following equation:
\begin{equation}
\label{eq1}
\frac{ \mu_{hkl} \ \lbrack \mu_B \rbrack}{1 \ \lbrack \mu_B \rbrack} = \frac{1}{f(\bm{Q}_{hkl})}\sqrt{ \frac{I^{hkl}_{obs}}{I^{002}_{obs}} / \frac{I^{hkl}_{cal}}{I^{002}_{cal}}},
\end{equation}
where $I^{hkl}_{obs}$ and $I^{002}_{obs}$ are the observed intensities of magnetic and 002 reflections, respectively. $I^{002}_{cal}$ is the intensity calculated from the crystal structure, and $I^{hkl}_{cal}$ is the intensity calculated by assuming the 1 $\mu_B$ magnetic moment is parallel to the $(hkl)$ plane. The information about the magnetic form factor $f(\bm{Q}_{hkl})$ of the $\pi^*$-electron of O$_2^-$ is also required, but this is not known experimentally. We derived the spatial distribution of the spin density of O$_2^-$ by DFT calculations and obtained $f(\bm{Q}_{hkl})$ by Fourier transform \cite{suppl}. The magnitudes of the ordered moments projected onto the (0~$\frac{1}{2}$~0) and (0~$\frac{1}{2}$~1) planes were then found to be $\mu_{0\frac{1}{2}0}=0.22(2)$ $\mu_B$ and $\mu_{0\frac{1}{2}1}=0.18(1)$ $\mu_B$, respectively. This means that they are both about the same size and are about 0.2 $\mu_B$.

The solid and dashed curves in Fig.~\ref{fig:magpeaks}(f) simulate the magnetic reflections assuming that the magnetic moments of 0.10 and 0.15 $\mu_B$ exist parallel to the (1~$\frac{1}{2}$~0) plane, respectively. The peak widths are set to the widths of the nearby nuclear reflections and the Gaussian functions are plotted. $f(Q_{1\frac{1}{2}0})$ is also considered.
From this result it is not known, within the limits of statistical accuracy, whether the projection of the ordered moment onto the (1~$\frac{1}{2}$~0) plane, namely $\mu_{1\frac{1}{2}0}$, is 0.10 $\mu_B$, but it cannot be greater than 0.15 $\mu_B$.
Based on all the above information, the direction of the ordered magnetic moment is examined. First, if the magnetic moment of 0.2 $\mu_B$ is oriented along the $c$-axis, then the observed values for $\mu_{0\frac{1}{2}0}$ and $\mu_{0\frac{1}{2}1}$ can be approximately explained, but 0.2 $\mu_B$ must also be observed as $\mu_{1\frac{1}{2}0}$. This assumption is therefore rejected. Second, if the magnetic moment of 0.2 $\mu_B$ is oriented along the $b$-axis, this is also rejected, since the 0~$\frac{1}{2}$~0 magnetic reflection must not be observed in this case. Third, if the magnetic moment of 0.2 $\mu_B$ is oriented along the $a$-axis, then $\mu_{0\frac{1}{2}0}=\mu_{0\frac{1}{2}1}=0.2$ $\mu_B$. This is in general agreement with the experimental results. In this case, $\mu_{1\frac{1}{2}0}$ is calculated to be 0.09 $\mu_B$, which is consistent with the result in Fig.~\ref{fig:magpeaks}(f). Therefore, we can conclude that the ordered moment is oriented along the $a$-axis. Given the magnetic anisotropy due to the dipole field, this is energetically advantageous since the orthorhombic lattice with $a<b$. Figure~\ref{fig:MagStrct} represents the most plausible antiferromagnetic structure of CsO$_2$.
This magnetic structure is also consistent with ESR studies \cite{ESR}, which report that the magnetic easy axis is perpendicular to the molecular axis.

The Curie-Weiss law of the magnetic susceptibility gives the effective magnetic moment $\mu_{eff}=2.05$ $\mu_B$ \cite{Miyajima_JPSJ}. This value can be reproduced with the g-value $g=2.37$, and the spin $s=1/2$, corresponding to $\mu=gs\mu_B=1.18$ $\mu_B$ as the magnitude of the localized magnetic moment. In addition, at the low temperature of 1.3 K, the magnetization process up to 600 kOe seems to saturate at about 1 $\mu_B$ \cite{Miyajima_JPSJ}. In comparison to these values, the value of the ordered moment obtained in this study, 0.2 $\mu_B$, is very small. It is expected that the ordered moment is suppressed by spin fluctuations arising from the one-dimensionality of the system.

In quasi-one-dimensional spin chains, the strong intrachain antiferromagnetic interaction $J$ gives rise to the Bonner-Fisher type magnetic susceptibility, while the weak interchain interaction $J'$ produces the long-range magnetic order at lower temperatures. Therefore, the ratio $T_N/J$ is an indicator of one-dimensionality. Quasi-one-dimensional systems with spin $s=1/2$ have been studied in many compounds containing the Cu$^{2+}$ ions. In KCuF$_3$, the ratio is $T_N/J=0.2$ and the ordered moment is reported to be $\mu_0=0.49(7)$ $\mu_B$ \cite{KCuF3}. In BaCu$_2$Si$_2$O$_7$, which is more one-dimensional, the ratio is $T_N/J = 0.03$ and the ordered moment is reported to be $\mu_0=0.16$ $\mu_B$ \cite{BaCu2Si2O7_1,BaCu2Si2O7_2}. CsO$_2$ has $J=42.8$ K and $T_N=9.6$ K \cite{Miyajima_JPSJ}, so $T_N/J=0.22$. This value is close to that of KCuF$_3$. On the other hand, the magnitude of the ordered moment is between the values of KCuF$_3$ and BaCu$_2$Si$_2$O$_7$. According to the chain-MF (mean field) model, Schulz gives the relation between $\mu_0$, $T_N$, $J$, and $J’$ \cite{Schulz}:
\begin{align}
|J'|=\frac{T_N}{1.28\sqrt{\ln(5.8J/T_N)}}\\
\mu_0=1.017\sqrt\frac{|J'|}{J}.
\end{align}
Applying this model to CsO$_2$ yields $|J’|=4.2$ K, $|J’|/J=0.097$, and $\mu_0=0.32$ $\mu_B$. The calculated $\mu_0$ value is not far from the value observed in this study. The smallness of $|J’|/J$ indicates the high one-dimensionality of this system. The cause is thought to be the $\pi^*$ orbital order.

We propose here a possible model of the orbital order in CsO$_2$ that satisfies the magnetic structure shown in Fig.~\ref{fig:MagStrct} as well as the condition of the exchange couplings, namely, $J>0$ (antiferromagnetic), $J’<0$ (ferromagnetic), and $J \gg |J’|$. For simplicity, we consider the $ab$-plane as a square lattice, although in reality, $a \ne b$, and the symmetry can be lowered even further. Figure~\ref{fig:orbital}(a) shows the possible model of the orbital order. The degeneracy of the $\pi^*$ orbitals is lifted due to the lowered symmetry and a ferro-orbital (antiferro-orbital) order is realized along the $b$-axis ($a$-axis). As shown in Fig.~\ref{fig:orbital}(b) and (c), the exchange coupling along the $b$-axis $J$ becomes antiferromagnetic because the electron transfer $t$ occurs only between the identical $\pi^*$ orbitals. On the other hand, as shown in Fig.~\ref{fig:orbital}(d), the exchange coupling along the $a$-axis $J’$ becomes ferromagnetic due to the orthogonalization of the neighboring orbitals. The electron transfer between the half-filled orbitals vanishes because they are orthogonalized, while the transfer $t’$ occurs between the orbitals that have the same orientation. The intra-molecule exchange coupling usually gives the ferromagnetic interaction. The electron transfer $t'$ is off-resonant, thus $|J'|$ is usually much smaller than $J$. This is consistent with the strong character of the antiferromagnetic chain along the $b$-axis. Such a relationship between molecular orbital order and spin structure has also been discussed in fullerene complexes, e. g. TDAE-C$_{60}$\cite{TDAEC60_1,TDAEC60_2,TDAEC60_3,TDAEC60_4,TDAEC60_5} and (NH$_3$)K$_3$C$_{60}$\cite{NH3K3C60_1,NH3K3C60_2}. It is also essentially the same as the Kanamori-Goodenough rule for the superexchange interaction in transition metal oxides.
We have discussed exchange interactions between the nearest neighbor molecules in the $ab$-plane. The second neighbor of a molecule is not in the same $ab$-plane, but there are eight molecules in the body diagonal direction of the unit cell. In the magnetic structure shown in Fig.~\ref{fig:MagStrct}, the interaction is frustrated because the molecule is ferromagnetically coupled to four of them and antiferromagnetically coupled to the other four of them. Therefore, only interactions in the $ab$-plane are expected to be valid.
Since this orbital order model is speculative, it will have to be proven experimentally in the future.

In conclusion, we performed a powder ND study on CsO$_2$. The 0~$\frac{1}{2}$~0 and 0~$\frac{1}{2}$~1 magnetic reflections were observed below $T_N$. An antiferromagnetic structure with the propagation vector of (0, $\frac{1}{2}$, 0) and the ordered moment aligned with the $a$-axis is proposed as the most plausible. The ordered moment is found to be as small as 0.2 $\mu_B$, which is attributed to the high one-dimensionality of the system. A possible model of orbital order that can explain the magnetic structure is proposed.

We acknowledge K.~Okada, S.~Harashima, K.~Ohoyama, K.~Iwasa, S.~Mori, Y.~Ohori, and T.~Shiomi
for their support, and H.~O.~Jeschke, K.~Shibata, J.~Otsuki, M.~Naka for fruitful discussions. The ND experiments were performed under the joint-research program (No. 21558 and No. 22594) of ISSP, the University of Tokyo.








\begin{thebibliography}{9}

\bibitem{Kugel}
K.~I.~Kugel and D.~I.~Khomskii, \href{https://doi.org/10.1070/PU1982v025n04ABEH004537}{Sov. Phys. Usp. \textbf{25}, 231 (1982).}

\bibitem{Tokura}
Y.~Tokura and N.~Nagaosa, \href{https://doi.org/10.1126/science.288.5465.462}{Science \textbf{288}, 462 (2000).}

\bibitem{NH3K3C60_2}
Y.~Iwasa and T.~Takenobu, \href{https://doi.org/10.1088/0953-8984/15/13/202}{J. Phys.: Condens. Matter \textbf{15}, R495 (2003).}

\bibitem{Zumsteg}
A.~Zumsteg, M.~Ziegler, W.~K\"{a}nzig, and M.~B\"{o}sch, \href{https://link.springer.com/article/10.1007/BF01458808}{Phys. Cond. Matter \textbf{17}, 267 (1974).}

\bibitem{Miyajima_PRB}
M.~Miyajima, F.~Astuti, T.~Fukuda, M.~Kodani, S.~Iida, S.~Asai, A.~Matsuo, T.~Masuda, K.~Kindo, T.~Hasegawa, T.~C.~Kobayashi, T.~Nakano, I.~Watanabe, and T.~Kambe, \href{https://doi.org/10.1103/PhysRevB.104.L140402}{Phys. Rev. B \textbf{104}, L140402 (2021).}

\bibitem{Miyajima_D}
M.~Miyajima, Doctoral Thesis, Okayama University (2021).

\bibitem{KO2_Neutron}
H.~G.~Smith, R.~M.~Nicklow, L.~J.~Raubenheimer, and M.~K.~Wilkinson, \href{https://doi.org/10.1063/1.1708328}{J. Appl. Phys. \textbf{37}, 1047 (1966).}

\bibitem{Fahmi_JPSJ}
F.~Astuti, M.~Miyajima, T.~Fukuda, M.~Kodani, T.~Nakano, T.~Kambe, and I.~Watanabe, \href{https://doi.org/10.7566/JPSJ.88.043701}{J. Phys. Soc. Jpn. \textbf{88}, 043701 (2019).}

\bibitem{Riyadi}
S.~Riyadi, B.~Zhang, R.~A.~de~Groot, A.~Caretta, P.~H.~M.~van~Loosdrecht, T.~T.~M.~Palstra, and G.~R.~Blake, \href{https://doi.org/10.1103/PhysRevLett.108.217206}{Phys. Rev. Lett., \textbf{108}, 217206 (2012).}

\bibitem{CsO2_TLL}
M.~Klanjsek, D.~Arcon, A.~Sans, P.~Adler, M.~Jansen, and C.~Felser, \href{https://doi.org/10.1103/PhysRevLett.115.057205}{Phys. Rev. Lett. \textbf{115}, 057205 (2015).}

\bibitem{Miyajima_JPSJ}
M.~Miyajima, F.~Astuti, T.~Kakuto, A.~Matsuo, D.~P.~Sari, R.~Asih, K.~Okunishi, T.~Nakano, Y.~Nozue, K.~Kindo, I.~Watanabe, and T.~Kambe, \href{https://doi.org/10.7566/JPSJ.87.063704}{J. Phys. Soc. Jpn. \textbf{87}, 063704 (2018).}


\bibitem{Fahmi_D}
F.~Astuti, Doctoral Thesis, Hokkaido University (2019).



\bibitem{VESTA}
K.~Momma and F.~Izumi, \href{https://doi.org/10.1107/S0021889811038970}{J. Appl. Cryst. \textbf{44}, 1272 (2011).}

\bibitem{suppl}
See the Supplemental Material at [\textit{URL will be inserted by publisher}] for details.

\bibitem{ESR}
M.~Labhart, D.~Raoux, W.~K\"{a}nzig, and M.~A.~B\"{o}sch, \href{https://doi.org/10.1103/PhysRevB.20.53}{Phys. Rev. B \textbf{20}, 53 (1979).}

\bibitem{KCuF3}
M.~T.~Hutchings, E.~J.~Samuelsen, G.~Shirane, and K.~Hirakawa, \href{https://doi.org/10.1103/PhysRev.188.919}{Phys. Rev. \textbf{188}, 919 (1969).}

\bibitem{BaCu2Si2O7_1}
I.~Tsukada, Y.~Sasago, K.~Uchinokura, A.~Zheludev, S.~Maslov, G.~Shirane, K.~Kakurai, and E.~Ressouche, \href{https://doi.org/10.1103/PhysRevB.60.6601}{Phys. Rev. B \textbf{60}, 6601 (1999).}

\bibitem{BaCu2Si2O7_2}
M.~Kenzelmann, A.~Zheludev, S.~Raymond, E.~Ressouche, T.~Masuda, P.~B\"{o}ni, K.~Kakurai, I.~Tsukada, K.~Uchinokura, and R.~Coldea, \href{https://doi.org/10.1103/PhysRevB.64.054422}{Phys. Rev. B \textbf{64}, 054422 (2001).}

\bibitem{Schulz}
H.~J.~Schulz, \href{https://doi.org/10.1103/PhysRevLett.77.2790}{Phys. Rev. Lett. \textbf{77}, 2790 (1996).}

\bibitem{TDAEC60_1}
T.~Kawamoto, \href{https://doi.org/10.1016/S0038-1098(96)00584-4}{Solid State Commun. \textbf{101}, 231 (1997).}

\bibitem{TDAEC60_2}
T.~Kawamoto, M.~Tokumoto, H.~Sakamoto, and K.~Mizoguchi, \href{https://doi.org/10.1143/JPSJ.70.1892}{J. Phys. Soc. Jpn. \textbf{70}, 1892 (2001).}

\bibitem{TDAEC60_3}
K.~Mizoguchi, M.~Machino, H.~Sakamoto, T.~Kawamoto, M.~Tokumoto, A.~Omerzu, and D.~Mihailovic, \href{https://doi.org/10.1103/PhysRevB.63.140417}{Phys. Rev. B \textbf{63}, 140417(R) (2001).}

\bibitem{TDAEC60_4}
T.~Kambe, K. ~Kajiyoshi, M.~Fujiwara, and K.~Oshima, \href{https://doi.org/10.1103/PhysRevLett.99.177205}{Phys. Rev. Lett., \textbf{99}, 177205 (2007).}

\bibitem{TDAEC60_5}
M.~Fujiwara, T.~Kambe, and K.~Oshima, \href{https://doi.org/10.1103/PhysRevB.71.174424}{Phys. Rev. B \textbf{71}, 174424 (2005).}

\bibitem{NH3K3C60_1}
H.~Tou, Y.~Maniwa, Y.~Iwasa, H.~Shimoda, and T.~Mitani, \href{https://doi.org/10.1103/PhysRevB.62.R775}{Phys. Rev. B \textbf{62}, R775 (2000).}


\end{thebibliography}
\end{document}